\documentclass[a4paper, conference]{IEEEtran}
\IEEEoverridecommandlockouts
\usepackage[nocompress]{cite}
\usepackage{amsmath,amssymb,amsfonts}
\usepackage{algorithmic}
\usepackage{graphicx}
\usepackage{textcomp}
\usepackage{xcolor}
\usepackage{amssymb}  
\usepackage{algorithm}
\begin{document}
    \title{\huge Study of Iterative Dynamic Channel Tracking for Multiple RIS-Assisted MIMO Systems\vspace{-0.5em} }

\author{Roberto C. G. Porto and Rodrigo C. de Lamare \vspace{-1.1em}

\thanks{The authors are with the Centre for Telecommunications Studies, Department of Electrical Engineering, Pontifical Catholic University of Rio de Janeiro. Emails: camara@aluno.puc-rio.br, delamare@puc-rio.br}}

\maketitle

    \begin{abstract}
The use of multiple Reconfigurable Intelligent Surfaces (RIS) has gained attention in 6G networks to enhance coverage. However, the feasibility of deploying multiple RIS relies on efficient channel estimation and reduced pilot overhead. To address these challenges, this work proposes an iterative channel estimation scheme that exploits low-density parity-check (LDPC) codes, channel coherence time, and iterative processing to improve estimation accuracy while minimizing pilot length. Encoded pilots are used to strengthen the iterative processing, leveraging both pilot and parity bits, while previous estimates are incorporated to further reduce overhead. Simulations consider a sub-6 GHz scenario with non-sparse channels and multiple RIS under both LOS and NLOS conditions. The results show that the proposed method outperforms existing approaches, achieving significant gains with substantially lower pilot overhead.
\end{abstract}


    \vspace{-1em}
\section{Introduction}
Reconfigurable Intelligent Surfaces (RIS) have recently emerged as a key technology for enhancing wireless communication capabilities in 6G networks. Multiple-RIS systems \cite{10767769} have been proposed to extend network coverage, enabling access point services over wider areas with low deployment and maintenance costs \cite{10439018}. Due to the versatility of these metasurfaces, there is a wide range of applications, including indoor, outdoor, and vehicular environments.

However, the practical deployment of multiple RIS faces significant challenges, particularly in channel estimation and pilot overhead reduction \cite{10818440}, which can be further degraded under time-varying \cite{9854102}. As the number of users and RIS elements increases, the pilot length required for reliable estimation may become prohibitive, representing a limiting factor for scalable system implementation.

%However, the practical deployment of multiple RIS faces significant challenges, particularly in channel estimation and pilot overhead reduction \cite{10818440}, which can be further degraded under time-varying conditions \cite{9854102}. Depending on the number of users and RIS elements, the required pilot length can become a limiting factor.

Prior works proposed pilot-aided estimation schemes \cite{9130088, 9839429}, sparse reconstruction methods \cite{10614235}, and prediction exploiting temporal correlation \cite{10818440}. Recent efforts such as \cite{10484981} have considered time-varying channels in RIS-assisted cell-free systems. Nonetheless, these methods rely on uncoded systems, where channel estimation is performed solely based on pilot symbols, without exploiting the parity bits available in encoded data to enhance estimation performance. Furthermore, some non-RIS studies investigated the use of iterative processing combined with coding to improve channel estimation. In \cite{4357052}, the authors proposed an iterative channel estimation scheme for OFDM systems using LDPC codes, leveraging Encoded Pilots (EP) to aid the iterative detection and decoding (IDD). Similarly, \cite{10319806} explored code-aided methods for channel estimation. To the best of our knowledge, there is no iterative estimation scheme designed for RIS-based systems due to the challenges associated with separating the channels for multiple users without significantly increasing the pilot overhead.

In this work, we propose a novel iterative channel estimation scheme for multi-RIS systems, exploiting temporal correlation and LDPC-coded pilots. Specifically, we develop an iterative detection, decoding, and estimation framework that encodes pilot symbols and leverages both the pilots and the parity bits to enhance channel estimation. Additionally, we exploit the temporal coherence of the channel by using previous estimates as inputs to the iterative estimator, increasing spectral efficiency and overall performance. A key aspect of our estimator is its generality, making it adaptable to various metasurface technologies, including active \cite{9998527}, beyond-diagonal \cite{10308579}, and holographic RIS \cite{10158690}. Numerical results demonstrate that the proposed scheme outperforms existing techniques under both line-of-sight (LOS) and non-line-of-sight (NLOS) conditions.

The remainder of this paper is structured as follows. Section II presents the proposed system model. In Section III, the proposed iterative channel estimation technique is outlined. Section IV discusses exploiting temporal correlation for efficient channel estimation. Section V presents the simulation results, and Section VI concludes the paper.

\textit{Notation:} Bold capital letters represent matrices, while bold lowercase letters denote vectors. The symbol $\mathbf{I_n}$ refers to an $n \times n$ identity matrix and $\text{diag}(\mathbf{A})$ is a diagonal matrix with the diagonal elements of $\mathbf{A}$. The sets of complex and real numbers are denoted by $\mathbb{C}$ and $\mathbb{R}$, respectively. [·]$^{-1}$, [·]$^{T}$, and [·]$^H$ denote the inverse, transpose, and conjugate transpose, respectively. The Kronecker product is represented by $\otimes$.

\begin{figure}
%\vspace{-3em}
    \includegraphics[width=0.50\textwidth]{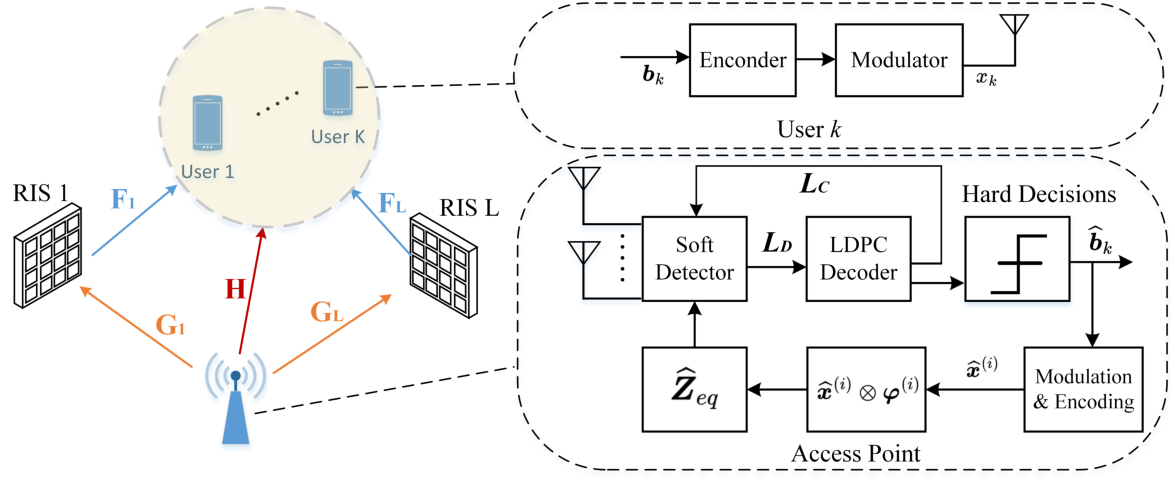}
    \vspace{-0.95em}
    \caption{System model of an IDD multiuser multiple-antenna system.}
    \label{fig:blockdiagram}
    \vspace{-1.2em}
\end{figure}

% \begin{figure*}
% %\vspace{-3em}
%     \centerline{\includegraphics[width=0.50\textwidth]{Figures/spawc_fig01.eps}}
%     \vspace{-0.95em}
%     \caption{System model of an IDD multiuser multiple-antenna system.}
%     \label{fig:blockdiagram}
%     \vspace{-1.2em}
% \end{figure*}
   \section{System Model}

A single-cell multiuser system featuring a multiple-antenna access point (AP) \cite{mmimo,wence} assisted by $L$ reflective intelligent surfaces (RIS) is considered, as illustrated in Fig. \ref{fig:blockdiagram}. In this configuration, the AP is equipped with $M$ antennas that support $K$ users, each equipped with a single antenna. The information bits for each user are encoded via individual LDPC channel encoders and subsequently modulated to $x_k$ using a QPSK modulation scheme. The transmit symbols $x_k$ have zero mean and share the same energy, with $E[|x_k|^2] = \sigma^2_x$. These modulated symbols are transmitted over block-fading channels.

Each RIS has $N$ reflecting elements, and their reflection coefficients are modeled as a complex vector $\boldsymbol{\varphi}_j \triangleq [e^{j\theta_1}, \dots, e^{j\theta_N}]^T$, where $\theta_n$ represents the phase shift of the $n$th unit of the RIS-$j$. The corresponding diagonal reflection matrix at time instant $i$ is given by and $\mathbf{\Phi}_j^{(i)} \triangleq \text{diag}(\boldsymbol{\varphi}_j)$. The received signal $\mathbf{y}^{(i)}$ at time instant $i$ can be expressed as
\begin{equation}
    \boldsymbol{y}^{(i)} = (\mathbf{H} + 
    \sum_{j=1}^L\mathbf{G}_j\mathbf{\Phi}_j^{(i)}\mathbf{F}_j +
     \mathbf{E}^{(i)})
    \boldsymbol{x}^{(i)} + \mathbf{n}^{(i)},
    \label{eq01}
\end{equation}
where $\mathbf{H} \in \mathbb{C}^{M \times K}$, $\mathbf{G}_j \in \mathbb{C}^{M \times N}$, and $\mathbf{F}_j \in \mathbb{C}^{N \times K}$ represent the communication links from the AP to the users, from the AP to RIS-$j$, and from RIS-$j$ to the users, respectively. The vector $\mathbf {x}^{(i)} \triangleq [x_1^{(i)}, \dots, x_K^{(i)}]^T$ represents the coded symbols transmitted by the users at time instant $i$, while and $\mathbf{n}^{(i)} \sim \mathcal{CN}(\mathbf{0}_M, \sigma_n^2\mathbf{I}_M)$ represents the noise. 

The signal reflected between different RIS at instant $i$ is represented by $\mathbf{E}^{(i)}$. Due to the multiplicative fading effect \cite{9998527}, this element has a negligible effect on the received signal and is ignored in the remainder of this work. Note that RIS can also be deployed to reduce the effect of $\mathbf{E}^{(i)}$.

The contributions of different RIS to the users can be grouped in a concise expression by
% \begin{equation}
%   \boldsymbol{y}^{(i)} = \left(\mathbf{H} + [\boldsymbol{G}_1 \dots \boldsymbol{G}_L]
% \begin{bmatrix}
% \boldsymbol{\Phi}_1^{(i)} & \boldsymbol{0} & \boldsymbol{0} \\
% \boldsymbol{0} & \ddots           & \boldsymbol{0} \\
% \boldsymbol{0} & \boldsymbol{0}   & \boldsymbol{\Phi}_L^{(i)} \\
% \end{bmatrix} 
% \begin{bmatrix}
% \boldsymbol{F}_1 \\
% \vdots \\
% \boldsymbol{F}_L
% \end{bmatrix}\right)\boldsymbol{x}^{(i)} + \boldsymbol{n}^{(i)},
%     \label{eq02}
% \end{equation}
\begin{align}
\boldsymbol{y}^{(i)} =&  [\mathbf{G}_1 \dots \mathbf{G}_L]
\begin{bmatrix}
\boldsymbol{\Phi}_1^{(i)} & \boldsymbol{0} & \boldsymbol{0} \\
\boldsymbol{0} & \ddots           & \boldsymbol{0} \\
\boldsymbol{0} & \boldsymbol{0}   & \boldsymbol{\Phi}_L^{(i)} \\
\end{bmatrix} 
\begin{bmatrix}
\mathbf{F}_1 \\
\vdots \\
\mathbf{F}_L
\end{bmatrix}\boldsymbol{x}^{(i)}
\\
&+ \mathbf{H}\boldsymbol{x}^{(i)} + \boldsymbol{n}^{(i)}. \nonumber 
\end{align}
Grouping the matrices of the communication links leads to
\begin{equation}
    \boldsymbol{y}^{(i)} = (\mathbf{H} + \mathbf{G}_p\boldsymbol{\Phi}_p^{(i)}\mathbf{F}_p)\boldsymbol{x}^{(i)} + \boldsymbol{n}^{(i)} = \mathbf{\bar{H}}_p^{(i)}\boldsymbol{x}^{(i)}  + \boldsymbol{n}^{(i)}, 
    \label{eq03}
\end{equation}
where $\mathbf{\bar{H}}_p$ represents the equivalent channel between the AP and the users. Equation (\ref{eq03}) is similar to that used for the representation of a single RIS-assisted MIMO system, which differs only in how the matrices are grouped. Therefore, the same techniques can be used here. In addition to the grouped RIS phase-shift matrix, $\boldsymbol{\Phi}_p^{(i)}$ remains a diagonal-only matrix. Since the matrix $\mathbf{\Phi}$ is diagonal, the received signal in (\ref{eq03}) can be written in terms of $\boldsymbol{\varphi}^{(i)}_p$ as given by 
\begin{equation}
     \boldsymbol{y}^{(i)} = \sum_{k=1}^K\boldsymbol{h}_kx_k^{(i)} + \sum_{k=1}^K\mathbf{Z}_k\boldsymbol{\varphi}_p^{(i)}x_k^{(i)} + 
     \mathbf{n}^{(i)},
    \label{eq:eq04}
\end{equation}
where $\mathbf{Z}_k = \mathbf{G}_p \text{diag}(\boldsymbol{f}_{p,k})$, with  $\boldsymbol{f}_{p,k}$  denoting the 
$k$th column of the matrix $\mathbf{F}_p$.

An estimate $\hat{x}_k$ of the transmitted symbol is obtained by applying a linear receive filter $\boldsymbol{w_k}$ to the received signal:
\begin{equation}
    \hat{x}_k^{(i)} = (\mathbf{w}_k^{(i)})^H\boldsymbol{y}^{(i)}.
    \label{detection_estimate_1}
\end{equation}
This symbol estimate uses successive interference cancellation (SIC), where the value of $\boldsymbol{\varphi}$ is fixed and $\mathbf{w_k}$ is chosen to minimize the mean-square error (MSE) between the transmitted symbol $x_k$ and the filter output.
{\begin{equation} 
    \mathbf {w}_{k}=\arg \min _{ \tilde{\mathbf{w}}_{k}} E\left [{\left \vert{ x_{k}-\tilde{\mathbf{w}}_{k}^{H}\mathbf {y}_k}\right \vert ^{2}}\right]. 
\end{equation}}
It can be shown that the solution is given by
\begin{equation}
    \mathbf{w}_k = \left(\frac{\sigma^2_n}{\sigma^2_x}\mathbf{I_{n_r}} + \mathbf{\bar{H}}_p \mathbf{\Delta}_k\mathbf{\bar{H}}_p^{\rm H} \right)^{-1}\boldsymbol{\bar{h}}_k,
    \label{eq:w}
\end{equation}
where $\mathbf {\bar{H}} \triangleq [\mathbf{\bar{h}_1}, \dots, \mathbf{\bar{h}_K}]^H$ is the equivalent channel and the covariance matrix $\mathbf{\Delta_k}$  is
{\begin{equation}
    \mathbf{\Delta}_k = \text{diag}\left[\frac{\sigma^2_{x_{1}}}{\sigma^2_x}\dots \frac{\sigma^2_{x_{k-1}}}{\sigma^2_x}, 1, \frac{\sigma^2_{x_{k+1}}}{\sigma^2_x},\dots,\frac{\sigma^2_{x^2_{K}}}{\sigma^2_x}  \right],
\end{equation}}
where $\sigma^2_{x_{i}}$ is the variance of the $k$th user, computed as:
{\begin{equation}
\sigma_{x_{k}}^{2}=\sum\limits_{x\in {\cal A}}\vert x-\bar{k} _{i}\vert ^{2}P(x_{k}=x). 
\end{equation}} 

The optimization of reflection parameters builds upon the work in \cite{10747209}, which focuses on the design of reflection parameters using the minimum mean-square error (MMSE) criterion. This approach yields effective results for refining LLRs within an IDD system. Specifically, an MMSE receiver filter was employed to facilitate SIC at the receiver. Alternatively, a designer can apply more sophisticated interference cancellation strategies \cite{jidf,spa,mfsic,dfcc,mbdf,bfidd,did,rrber,listmtc,msgamp,msgamp2,comp,vrce,llrap,iddocl}.

% \subsection{Time-varying Channel Model}

The flat-fading MIMO channel is modeled as a discrete-time first-order Gauss-Markov process. In this model, the channel remains constant within each transmission block and exhibits time correlation across successive blocks, which reflects practical scenarios where the block duration is smaller than or comparable to the coherence time of the wireless channel~\cite{10484981}.

For simplicity, we assume that all users are equidistant from the access point (AP), allowing the channel evolution between adjacent blocks to be modeled as:
\begin{equation}
    \mathbf{H}^{(b)} = \rho \mathbf{H}^{(b-1)} + \sqrt{1-\rho^2}\, \mathbf{W}^{(b)}_{\text{age}},
    \label{eq:markov}
\end{equation}
where $\rho \in [0,1]$ denotes the temporal correlation coefficient, $\mathbf{W}^{(b)}_{\text{age}} \sim \mathcal{CN}(0,\sigma^2_{\text{age}}\mathbf{I})$ represents the noise introduced at each block, and $b$ denotes the block index. In this context, $\mathbf{H}^{(b-1)}$ serves as a temporally outdated yet correlated representation of the current channel.

    \section{Proposed Channel Estimation}
\label{sec:proposed}
% In this section, we propose a novel channel estimation technique for multi-RIS-assisted MIMO systems that leverages iterative detection and decoding with EP and LDPC decoding to reduce the number of pilots required to estimate the cascaded RIS coefficients. All the pilot symbols are encoded with the data bits using a systematic encoder; therefore, the pilots remain unaltered and can be considered known by the receiver, as illustrated in Fig. \ref{fig:package}. 
% The technique begins by estimating the direct channel and obtaining an initial raw estimation of the reflected RIS channel. After the first iteration, the entire estimated symbol packet is used as pilots to refine the estimation of the reflected channel. Since the receiver is prone to incorrect symbol estimates, this procedure is performed iteratively to refine the channel estimation in each iteration.

\begin{figure}
%\vspace{-3em}
\centerline{\includegraphics[width=0.45\textwidth]{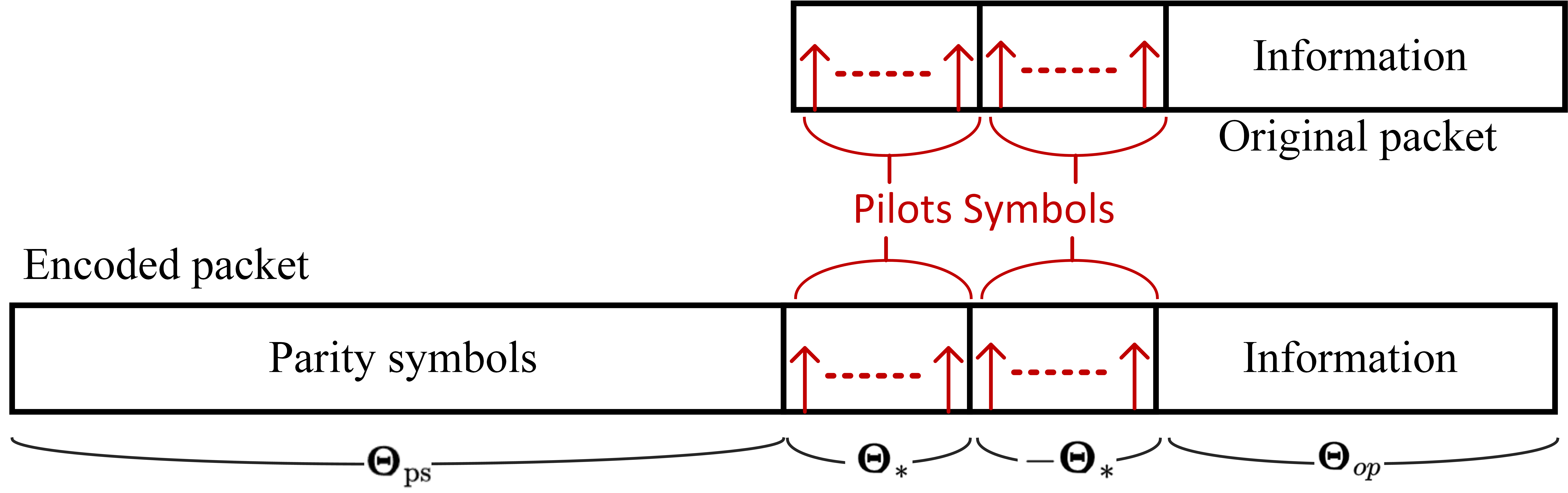}}
    \vspace{-0.725em}
    \caption{Systematically encoded packet and original post-modulation packet.}
    \label{fig:package}
    \vspace{-1em}
\end{figure}

% To introduce this approach, we first explain how the direct channel can be estimated, then describe the coarse estimation of the reflected channel, and eventually show how EP and the iterative processing enhance the overall estimation accuracy.
This section proposes a novel channel estimation method for multi-RIS MIMO systems using iterative detection and decoding with EP and LDPC, aiming to reduce pilot overhead. Pilot symbols are embedded with data via a systematic encoder and remain known to the receiver (Fig.~\ref{fig:package}). The process starts with direct channel estimation and a coarse estimate of the reflected channel. In subsequent iterations, the entire estimated symbol block is reused as pilots to refine the RIS-related coefficients. Due to possible detection errors, the refinement is performed iteratively. We begin by detailing the direct channel estimation, followed by the coarse RIS estimation, and then the use of EP for iterative refinement.

\subsection{Direct Channel Estimation}
\label{subsec:direct}
To estimate the direct channel, we used an always-on channel estimation approach without switching off the selected RIS elements \cite{9839429}. This protocol involves dividing the pilot sequence into two partitions of equal size, allowing us to represent the received signal for each partition as
\begin{equation}
    \boldsymbol{y}^{(j)} = (\mathbf{H} + \mathbf{G}_p\text{diag}(\boldsymbol{\varphi}_p^{(j)})\mathbf{F}_p)\boldsymbol{x}^{(j)} + \boldsymbol{n}^{(j)},
    \label{eq:partA}
\end{equation}
where $j$ belongs to the set of pilot symbols. We define the first and second partitions, respectively, as:  
\begin{equation}
\mathcal{P}_1 = \left\{ j \mid t \leq j \leq t + \frac{N_p}{2}-1\right\}    
\label{eq:partition01}
\end{equation}
\begin{equation}
\mathcal{P}_2 = \left\{ j \mid t + \frac{N_p}{2} \leq j \leq t + N_p -1\right\}    
\label{eq:partition02}
\end{equation}
where  $N_p$  denotes the total number of pilot symbols, and $t$  represents the index of the first pilot symbol.
By selecting the pilots and reflection parameter values such that $x^{(j)} = x^{\left(j + \frac{N_p}{2}\right)}$  and $\boldsymbol{\varphi}_p^{(j)} = -\boldsymbol{\varphi}_p^{\left(j + \frac{N_p}{2}\right)}$, we can define the sum and subtraction of each received signal as
\begin{equation}
  \frac{\boldsymbol{y}^{(j)}+\boldsymbol{y}^{(j+\frac{Np}{2})}}{2} = \mathbf{H}\boldsymbol{x}^{(j)} + \boldsymbol{w}
    \label{eq:directchannel}
\end{equation}
%and
\begin{equation}
    \frac{\boldsymbol{y}^{(j)}-\boldsymbol{y}^{(j+\frac{Np}{2})}}{2} = \textbf{G}_p\text{diag}(\boldsymbol{\varphi}_p^{(j)})\textbf{F}_p\boldsymbol{x}^{(j)} + \boldsymbol{w},
    \label{eq:reflectedchannel}
\end{equation}
where $\boldsymbol{w} \sim \mathcal{CN}(\mathbf{0_M}, \frac{\sigma_n^2}{2} \mathbf{I_M})$.
From (\ref{eq:directchannel}), we can apply conventional channel estimation methods to estimate only the direct channel. In this work, we use the linear MMSE (LMMSE) channel estimator given by 
\begin{equation}
    \hat{\mathbf{H}}_\text{LMMSE} = \mathbf{Y}_p\left(\mathbf{P}^H\mathbf{R}_\text{H}\mathbf{P}+\frac{\sigma_n^2}{2\sigma_x^2}\mathbf{I}\right)^{-1}\mathbf{P}^H\mathbf{R}_\text{H},
    \label{eq:lmmse}
\end{equation}
where $\mathbf{R}_\text{H} := E[\mathbf{H}\mathbf{H}^H]$ is the channel covariance matrix, $\mathbf{P}$ is the matrix representing the vectors of pilot symbols, and  $\mathbf{Y}_p$ is the received matrix obtained by summing the received signals from (\ref{eq:directchannel}). For this estimation, since both partitions are used to obtain the direct channel, a total of  $N_p = 2K$ pilot symbols are required.

\subsection{Cascate RIS Channel Estimation}
\label{subsec:cascate}
To estimate the coefficients of the reflected channel without interference from the direct link, we use (\ref{eq:reflectedchannel}). By rewriting the received signal in terms of $\boldsymbol{\varphi}_p^{(i)}$ using (\ref{eq:eq04}), the received signal from the RIS can be expressed as
\begin{equation} \boldsymbol{y}^{(j)}_\text{cascate}=\frac{\boldsymbol{y}^{(j)}-\boldsymbol{y}^{(j+\frac{Np}{2})}}{2} = \sum_{k=1}^K\mathbf{Z}_k\boldsymbol{\varphi}_p^{(j)}x_k^{(j)} + \boldsymbol{w}^{(j)}.
    \label{eq:cascate_1}
\end{equation}
Eq. \eqref{eq:cascate_1} can be rearranged by concatenating the matrices $\mathbf{Z}_k$ and eliminating the summation, resulting in 
% begin{equation}
%     \boldsymbol{y}^{(j)}_\text{cascate}= 
%     \begin{bmatrix}
%         \boldsymbol{Z}_1& \dots & \boldsymbol{Z}_K \\
%     \end{bmatrix}(\boldsymbol{x}^{(j)}\otimes{\varphi}_p^{(j)} ) = \boldsymbol{Z_\text{all}}\boldsymbol{\lambda}^{(j)},  
%     \label{eq:cascate_2}
% \end{equation}
\begin{align}
    \boldsymbol{y}^{(j)}_\text{cascate}&= [\mathbf{Z}_1, \dots, \mathbf{Z}_K](\boldsymbol{x}^{(j)}\otimes{\varphi}_p^{(j)} ) + \boldsymbol{w}^{(j)}
    \\
    &= \mathbf{Z_\text{all}}\boldsymbol{\lambda}^{(j)} + \boldsymbol{w}^{(j)} \nonumber 
\end{align}
where $\mathbf{Z_\text{all}} \in \mathbb{C}^{M \times KN}$ is the concatenated matrix of the cascaded channel for each user and $\boldsymbol{\lambda}^{(j)}=(\boldsymbol{x}^{(j)}\otimes{\varphi}_p^{(j)} )$ is a complex vector with KN elements.

Since the symbols of partition $\mathcal{P}_1$ are known by the receiver, we can assume that $\boldsymbol{\lambda}^{(j)}$ is also fully known. After $T$ time slots of pilot transmission, we can obtain the $M \times T$ overall measurement matrix $\mathbf{Y}_\text{cascate}=[\boldsymbol{y}_\text{cascate}^{(1)}, \dots,\boldsymbol{y}_\text{cascate}^{(T)}]$ as 
\begin{equation}
    \mathbf{Y}_\text{cascate}= 
    \mathbf{Z_\text{all}}\boldsymbol{\Lambda} + \mathbf{W},  
    \label{eq:cascate_2}
\end{equation}
where $\boldsymbol{\Lambda} = [\boldsymbol{\lambda}^{(1)}, \dots, \boldsymbol{\lambda}^{(T)}]$ and $\mathbf{W} = [\boldsymbol{w}^{(1)}, \dots, \boldsymbol{w}^{(T)}]$.
This expression can be considered equivalent to a standard channel estimation problem, which allows us to apply (\ref{eq:lmmse}) to estimate the coefficients. 

To mitigate multiuser interference, the matrix $\boldsymbol{\Lambda}$ should be orthogonal or semi-orthogonal, ensuring that the coefficients of $\boldsymbol{Z_\text{all}}$ can be estimated independently for each user. This matrix should be well conditioned to prevent noise amplification during equation inversion. To this end, we generate $\boldsymbol{\lambda}^{(i)}$ using Hadamard sequences for $\boldsymbol{x}^{(i)}$, while $\boldsymbol{\varphi}_p^{(i)}$ is derived from the DFT matrix \cite{9130088}. Note that to estimate all channel coefficients, $\boldsymbol{\Lambda}$ must have at least a rank of $KN$, which may be infeasible in some scenarios. Therefore, we suggest first obtaining a coarse approximation of the cascaded RIS channel using only a few pilots (number of pilots $\ll KN$), and then refining the estimates through iterative channel estimation.
\vspace{-0.5em}
\subsection{Proposed Iterative Channel Estimation}
\label{subsec:iterative}
In this step, our goal is to use the entire decoded symbol sequence as input, even if some symbols are incorrectly decoded. As shown in Fig. \ref{fig:blockdiagram}, the output of the IDD provides the decoded bits. We employ the encoded pilot scheme, which combines pilots with information bits using a systematic encoder. This enables iterative channel estimation and decoding, where the EPs contribute to both processes, enhancing overall performance \cite{4357052}. For iterative channel estimation, the first step is to reapply coding and modulation to transform the decoded bits back into symbols. Then, we compute the Kronecker product $(\boldsymbol{\hat{x}}^{(j)}\otimes{\varphi}_p^{(j)} )$ to derive $\boldsymbol{\hat{\Lambda}}$, which allows the application of standard channel estimation techniques to (\ref{eq:cascate_2}). However, it is crucial for $\boldsymbol{\hat{\Lambda}}$ to remain semi-orthogonal and well-conditioned. Since the phase configurations satisfying these conditions differ from the optimal ones derived in \cite{10747209}, we adopt a suboptimal design for the parity bits.

Assuming that the packet in Fig. \ref{fig:package} consists of $N_\mathrm{ps}$ parity symbols, $N_\mathrm{p}$ pilot symbols and $N_\mathrm{info}$ information symbols, and that $\boldsymbol{\Theta}_\mathrm{i} = [\boldsymbol{\varphi}^{(i)} \boldsymbol{\varphi}^{(i+1)} \dots]$ represents the concatenation of the reflection parameter vectors, we can express the reflection parameter vectors for each symbol as follows:
\begin{equation}
    \begin{matrix}  
    {\boldsymbol{\Theta}}_{\mathrm{ps}} =\left [{
\begin{array}{cccc} 
1 &1  &\cdots &1 \\ 
1 &\omega  &\cdots &\omega ^{N_\mathrm{ps}-1} \\ 
\vdots  &\vdots &\ddots &\vdots \\ 
1 &\omega ^{N-1}  &\cdots &\omega ^{(N-1)(N_\mathrm{ps}-1)} \end{array}}\right]\end{matrix},
\end{equation}
\vspace{-0.75em}
\begin{equation}
    \boldsymbol{\Theta}_\mathrm{p} = [\boldsymbol{\Theta}_\mathrm{*} -\boldsymbol{\Theta}_\mathrm{*}],
\end{equation}
\vspace{-0.75em}
\begin{equation}
    \begin{matrix}  
    {\boldsymbol{\Theta}}_{*} =\left [{
\begin{array}{cccc} 
1 &1  &\cdots &1 \\ 
1 &\varpi  &\cdots &\varpi ^{N_\mathrm{p}/2-1} \\ 
\vdots  &\vdots &\ddots &\vdots \\ 
1 &\varpi ^{N-1}  &\cdots &\varpi ^{(N-1)(N_\mathrm{p}/2-1)} \end{array}}\right]\end{matrix},
\end{equation}
\vspace{-0.75em}
\begin{equation}
    \boldsymbol{\Theta}_{\mathrm{o}} = [\boldsymbol{\varphi}_\mathrm{o} 
    \dots
    \boldsymbol{\varphi}_\mathrm{o}], \quad
    \boldsymbol{\varphi}_\mathrm{o} \text{: optimal \cite{10747209}}
\end{equation}
where $\omega = e^\frac{-2\pi i}{N_\mathrm{ps}}$ and $\varpi = e^\frac{-4\pi i}{N_\mathrm{p}}$. If $N<\frac{N_\mathrm{p}}{2}$, to ensure pseudo-orthogonality between the RIS elements, we concatenate $N \times N$ DFT matrices such that the condition $uN \geq \frac{N_\mathrm{p}}{2}$ is met, where $u \in \mathbb{N}^*$ is the number of concatenated matrices. 
This ensures pseudo-orthogonality in $\boldsymbol{\Lambda}$ for both sequences of parity bits and pilots, while preserving the information symbols with the optimal reflection parameters. Note that since the same optimal phase configuration is applied to the information symbols, $\boldsymbol{\Theta}_{\mathrm{o}}$ is a low-rank matrix, which minimally contributes to the channel estimation.

% Subsequently, we obtain the Kroneker product $(\boldsymbol{x}^{(j)}\otimes{\varphi}_p^{(j)} )$ to obtain $\Lambda$, and can use the conventional mimo channel estimation techniques to (\ref{eq:cascate_2}). However, as mentioned previously, it is important that $\boldsymbol{\Lambda}$ be semi-orthogonal and well conditioned. Because the phases that enable these characteristics are not the ones obtained based on (\ref{eq05}), we adopt a non-optimal approach for the parity bits. For the pilots, as previously stated, we used a DFT matrix, whereas for the data symbols, we applied a metric based on (\ref{eq05}). This strategy ensures pseudo-orthogonality in $\boldsymbol{\Lambda}$ for parity bits and pilots.

The initial symbol estimates may contain errors, which can propagate to the channel estimation. To address this, the channel estimates are iteratively refined and the symbols are reprocessed until convergence is achieved or a predefined maximum number of iterations is reached. The pseudo-code of the proposed algorithm is shown below.

% \begin{algorithm}[H]
% \footnotesize
%     \label{algor1}
%     \begin{algorithmic}[1]
%         \caption{Proposed Iterative Channel Estimation}\label{alg:cap}
%         \STATE Users transmit pilot symbols using Hadamard sequences for $\boldsymbol{x}^{(i)}$, while $\boldsymbol{\varphi}_p^{(i)}$ is derived from the DFT matrix.
%         \STATE The BS removes the reflected signal based on (\ref{eq:directchannel}) and estimates $\mathbf{H}$ using (\ref{eq:lmmse}).
%         \STATE The BS removes the direct signal based on (\ref{eq:reflectedchannel}) and performs a coarse estimation of $\boldsymbol{Z_\text{all}}$ using (\ref{eq:cascate_2}) and (\ref{eq:lmmse}).
%         \\
%         \textbf{Iterative Estimation Process}
%         \FOR{$t=1$ to Maximum Number of Iterations}
%             \STATE \textbf{IDD Scheme - SIC}: performed based on (\ref{eq:ldlc}).
%             \STATE Using the LDPC decoder outputs, refine the channel estimation based on (\ref{eq:lmmse}), incorporating all symbols in the packet (both parity and data).
%             \IF{NMSE of estimated channels $<$ tolerance value}
%                 \STATE \textbf{break}
%             \ENDIF
%         \ENDFOR
%     \end{algorithmic}
% \end{algorithm}
% \vspace{-0.5cm}
\begin{algorithm}[H]
\footnotesize
    \label{algor1}
    \begin{algorithmic}[1]
        \caption{Proposed Iterative Channel Estimation}\label{alg:cap}
        \STATE Estimate the direct channel ${\mathbf{H}}$ using (\ref{eq:lmmse}) based on (\ref{eq:directchannel}).
        \STATE Perform initial estimation of the reflected channel ${\boldsymbol{Z_\text{all}}}$ using (\ref{eq:lmmse}) based on (\ref{eq:reflectedchannel}).
        \FOR{$t = 1$ to Max Iterations or until $\text{tol} > \text{NMSE}$}
            \STATE \textbf{IDD Scheme - SIC}: Apply the scheme as defined in Sec. \ref{sec:temporal}.
            \STATE Refine the reflected channel estimation $\hat{\boldsymbol{Z}}_\text{all}$ using (\ref{eq:cascate_2}) and the estimator (\ref{eq:lmmse}).
        \ENDFOR
    \end{algorithmic}
\end{algorithm}
\vspace{-0.5cm}

% \begin{table}[!ht]
% \caption{Effective Code Rates for the Same Packet Size $N$ and Code Rate $R=1/2$}
%   \label{tab:1} \vspace{-0.2cm}
%   \centering
% \begin{tabular}{|c|c|c|}
% \hline
%  & Encoded Pilot  & Reference  \\ \hline
% Nº of pilots symbols  & $N_p$ & $N_p$ \\[3pt] \hline
% Nº of data symbols  & $\frac{N-N_p}{2}$ & $\frac{N-2N_p}{2}$  \\[3pt] \hline
% Effective code Rate & $0.5 - \frac{N_p}{2N}$ & $0.5 - \frac{N_p}{N}$ \\[3pt] \hline
% \end{tabular}
% \vspace{0.00cm}
% \end{table}

    %\subsection{Proposed Channel Tracking Protocol}
\section{Exploiting Temporal Correlation for Efficient Channel Estimation}
\label{sec:temporal}

In this section, we propose exploiting the temporal correlation between adjacent transmitted blocks to improve the efficiency of the proposed iterative channel estimation algorithm. Assuming that the enhanced channel estimate from the previous time instant, obtained through the iterative algorithm, is available, the current reflected channel can be estimated as:
\begin{equation}
    \hat{\mathbf{Z}}_\text{all}^{(b)} = \mathbf{Z}_\text{all}^{(b-1)} + \mathbf{E}^{(b-1)}_{\rm{est}}
\end{equation}
where $\mathbf{E}^{(b-1)}_{\rm{est}}$  denotes the estimation error associated with the previous block (with block index $b-1$). Under the assumption of normalized channels with unit average power and using the time-varying model (\ref{eq:markov}), the normalized mean squared error (NMSE) associated with this estimate can be expressed as:
\begin{equation} 
    \text{NMSE}_{\text{last}} = \frac{|| {\mathbf{Z}}_\text{all}^{(b)} - \hat{\mathbf{Z}}_\text{all}^{(b)} ||_F^2}{||{\mathbf{Z}}_\text{all}^{(b)} ||_F^2} \approx 2(1 - \rho) + \sigma_{\rm{est}}^2, 
\end{equation}
where $\sigma_{\rm{est}}^2 =  \frac{1}{MK}\mathbb{E}[||{\mathbf{E}}^{(b-1)}_{\rm{est}}||^2_F]$ represents the residual error after iterative estimation.

This enables a direct comparison with the coarse, low-rank estimate introduced in subsection \ref{subsec:cascate}. In particular, if the inequality
\begin{equation} 
2(1 - \rho) + \sigma_e^2 < \text{NMSE}_{\text{coarse}} 
\label{eq:comparison}
\end{equation}
is satisfied, then reusing the enhanced estimate from the previous frame offers a more accurate initialization for the iterative algorithm than relying solely on a fresh, structurally constrained raw estimate.

This temporal reuse strategy not only improves accuracy but can also reduce the number of iterations required for the NMSE to converge to that of a full-pilot benchmark, thereby improving convergence speed and efficiency.

Alternatively, one may prioritize reducing pilot overhead in subsequent frames by relying on coarser updates, acknowledging that this may increase $\sigma_e^2$ and lead to a moderate performance loss. To mitigate the accumulation of residual error, the channel estimation can be scheduled periodically, for instance, every $N_f$ frames.

\begin{figure*}
%\vspace{-3em}
    \centerline{\includegraphics[width=0.80\textwidth]{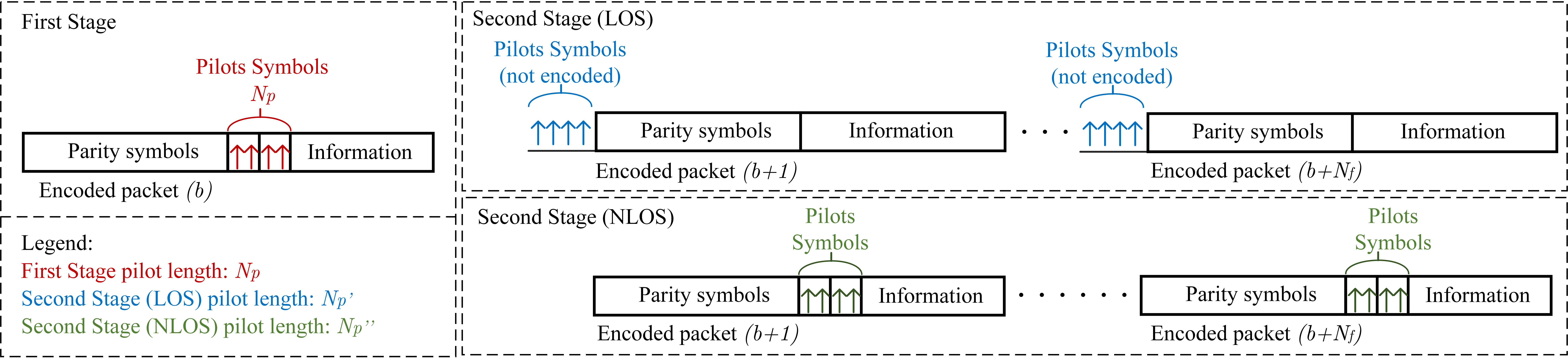}}
    \vspace{-0.75em}
    \caption{Proposed Channel Tracking Protocol for LOS and NLOS scenarios.}
    \label{fig:protocol}
    \vspace{-1.2em}
\end{figure*}

We propose a two-stage channel tracking protocol for both LOS and NLOS scenarios. In the \textbf{First Stage}, the steps from Section~\ref{sec:proposed} are executed, treating channel estimation as an iterative process refined over time.
In the \textbf{Second Stage}, temporal coherence is exploited and divided into two cases, as shown in Fig.~\ref{fig:protocol}:

\begin{itemize} 
    \item \textbf{LOS:} Only the cascaded channel is iteratively estimated; pilot symbols for the direct channel are still needed but remain uncoded. Specifically, $N_p'$ pilots are sent before the frame block, and the direct channel is estimated using the LMMSE estimator in (\ref{eq:lmmse}). For the reflected channel, the prior estimate is reused ($\mathbf{\hat{Z}}\text{all}^{(b)} = \mathbf{\hat{Z}}\text{all}^{(b-1)}$), followed by the iterative steps from Section~\ref{subsec:iterative}.
    \item \textbf{NLOS:} With no direct link, the always-on direct channel protocol from Section~\ref{subsec:direct} is unnecessary. A reduced number of pilots, $N_p'' \ll N_p$, is used. As in LOS, the previous estimate is reused ($\mathbf{\hat{Z}}_\text{all}^{(b)} = \mathbf{\hat{Z}}_\text{all}^{(b-1)}$), and $N_p''$ is retained solely for SIC during decoding.
\end{itemize}
% We propose a two-stage channel tracking protocol designed to accommodate both line-of-sight (LOS) and non-line-of-sight (NLOS) scenarios. In the \textbf{First Stage}, the steps outlined in Section~\ref{sec:proposed} are executed, where channel estimation is treated as an iterative process to refine the estimates over time.
% In the \textbf{Second Stage}, the temporal coherence of the channel is exploited. This stage is divided into two parts shown in Fig.~\ref{fig:protocol}:

% \begin{itemize}
%     \item \textbf{LOS Scenario:} Since only the cascaded channel is iteratively estimated, pilot symbols for the direct channel must still be transmitted. However, these pilots are not encoded. Specifically, $N_p'$ pilot symbols are transmitted ahead of the frame block, and the direct channel is estimated using the LMMSE estimator in (\ref{eq:lmmse}). For the reflected (cascaded) channel, the previous estimate is reused
%         ($\mathbf{\hat{Z}}_\text{all}^{(b)} = \mathbf{\hat{Z}}_\text{all}^{(b-1)}$),
%     and the iterative estimation steps in Section~\ref{subsec:iterative} are then followed.

%     \item \textbf{NLOS Scenario:} In the absence of a direct link, the always-on direct channel protocol described in Section~\ref{subsec:direct} is not required. Instead, a reduced number of pilot symbols, $N_p'' \ll N_p$, is employed. As in the LOS case, the previous channel estimate is reused
%         ($\mathbf{\hat{Z}}_\text{all}^{(b)} = \mathbf{\hat{Z}}_\text{all}^{(b-1)}$).
%     The pilot symbols $N_p''$ are preserved solely to support SIC during decoding.
% \end{itemize}

     \section{Numerical Results}

A short-length regular LDPC code \cite{ldpc,memd} with a block length of $N=512$ and rate $R = 1/2$ was considered with QPSK modulation. A sum-product algorithm was used for decoding although more sophisticated decoders can be adopted \cite{vfap, msgamp,msgamp2} The channel is assumed to experience block fading and the estimation of the channel state information (CSI) is done at the receiver. Two uplink scenarios were evaluated: LOS ( weakened direct link with strong reflected link) and NLOS (no direct link and strong reflected link). The path loss models follow the 3GPP standard \cite{access2010further}, and the system parameters are presented in Table \ref{tab:parameters}. To account for small-scale fading effects, a Rayleigh fading channel model was adopted for all channels. \vspace{-0.75em}

% \begin{table}[ht]
%     \caption{Simulation Parameters}\vspace{-0.75em}
%     \label{tab:parameters}
% \centering
% \begin{tabular}{c|c}
% \hline
% {\textbf{Parameters}}               & { \textbf{Values}} \\ \hline
% Frequency                           & 5 GHz              \\ \hline
% Bandwidth                           & 1 MHz              \\ \hline
% Sampling Frequency                  & 4 MHz               \\ \hline
% Noise power spectral density        & -170 dBm/Hz              \\ \hline
% Path loss AP-RIS; RIS-Users (dB)    & $37.3+22log_{10}(d)$  \\ \hline
% Path loss AP-Users (dB)             & $32.4+30log_{10}(d)$  \\ \hline
% Numer of AP antennas                & 8       \\ \hline
% Number of Users                     & 4       \\ \hline
% Number of RISs                      & 2      \\ \hline
% Number of Cells (per RIS)           & 16     \\ \hline
% Location of AP                      & (0 m, 0 m, 0 m)       \\ \hline
% Location of RIS$_1$                 & (500 m, 10 m, 0 m)     \\ \hline
% Location of RIS$_2$                 & (500 m, -10 m, 0 m)    \\ \hline
% Geometric center of users positions & (500 m, 0 m, 0 m)      \\ \hline
% Users Spatial Radius                & 5 m                   \\ \hline
% Doppler Frequency  (LOS scenario)             & 150 Hz                   \\ \hline
% Doppler Frequency  (NLOS scenario)             & 100 Hz                   \\ \hline
% \end{tabular}
% \end{table}

\begin{table}[ht]
    \caption{Simulation Parameters}
    \vspace{-1em}
    \label{tab:parameters}
    \centering
    \scriptsize
    \renewcommand{\arraystretch}{0.95}
    \begin{tabular}{c|c}
        \hline
        \textbf{Parameters}               & \textbf{Values} \\ \hline
        Frequency                         & 5 GHz           \\ \hline
        Bandwidth                         & 1 MHz           \\ \hline
        Sampling Frequency                & 4 MHz           \\ \hline
        Noise PSD                         & -170 dBm/Hz     \\ \hline
        Path Loss AP-RIS; RIS-Users       & $37.3+22\log_{10}(d)$ dB \\ \hline
        Path Loss AP-Users                & $32.4+30\log_{10}(d)$ dB \\ \hline
        Number of AP Antennas             & 8               \\ \hline
        Number of Users                   & 4               \\ \hline
        Number of RISs                    & 2               \\ \hline
        Number of Cells (per RIS)         & 16              \\ \hline
        AP Location                       & (0, 0, 0) m      \\ \hline
        RIS$_1$ Location                  & (500, 10, 0) m   \\ \hline
        RIS$_2$ Location                  & (500, -10, 0) m  \\ \hline
        Users' Geometric Center           & (500, 0, 0) m    \\ \hline
        User Spatial Radius               & 5 m             \\ \hline
        Doppler Freq. (LOS)               & 150 Hz          \\ \hline
        Doppler Freq. (NLOS)              & 100 Hz          \\ \hline
    \end{tabular}
\end{table}

\begin{figure}
    \vspace{-1em}
\includegraphics[width=0.45\textwidth]{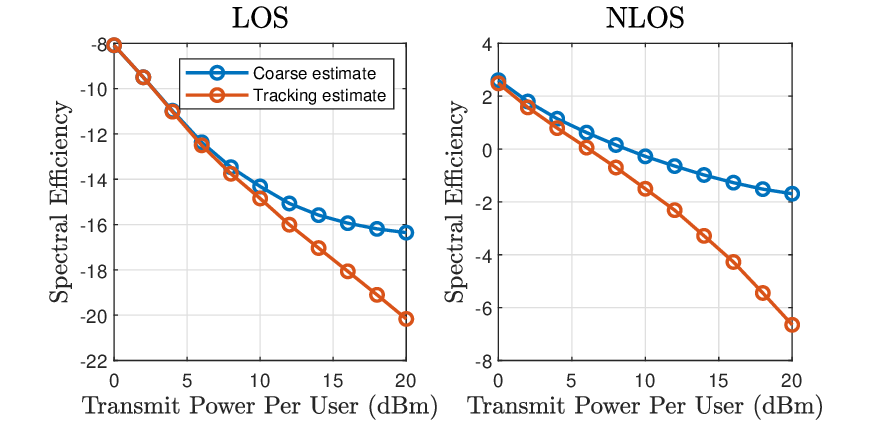}
  \vspace{-1em}
    \caption{Channel estimates comparison.}
    \label{fig:comparison}
    \vspace{-0.5em}
\end{figure}

\begin{figure}
    \vspace{-1em}
\includegraphics[width=0.425\textwidth,height=5.0cm]{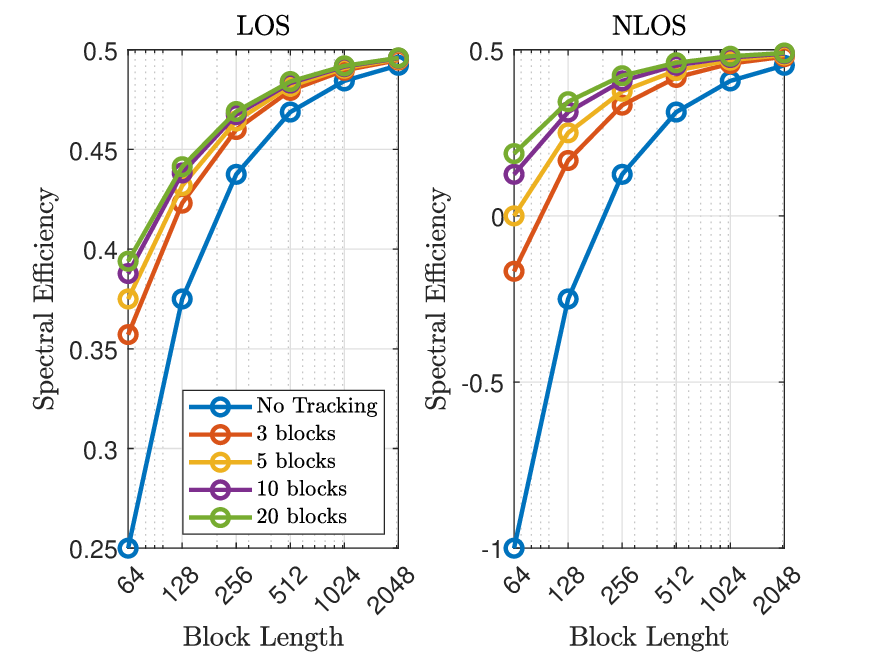}    \vspace{-1.0em}
    \caption{Spectral efficiency comparison. For the LOS scenario,  $N_p = N_p' = 16$; for the NLOS scenario, $N_p = 96$  and  $N_p'' = 16$.}
    \label{fig:efficiency}
\vspace{-0.75em}
\end{figure}

For both scenarios, the AP and RIS are static, while users are mobile. In the LOS case, a Doppler frequency of \(150\,\mathrm{Hz}\) yields a user speed \(v_d = c \frac{f_{\text{Doppler}}}{f_{\text{system}}} \approx 30\,\mathrm{m/s}\), with \(c\) as the speed of light. The coherence time is \(t_c \approx 0.423/f_{\text{Doppler}} \approx 2.28\,\mathrm{ms}\). Assuming two samples per symbol, the block duration is \(128\,\mu\mathrm{s} \ll 2.28\,\mathrm{ms}\), justifying the constant-channel assumption per block. The inter-block temporal correlation is then \(\rho \approx 0.9990\).

% For both scenarios, it is assumed that the AP and RIS are static, while the users exhibit relative motion. In the LOS case, we consider a Doppler frequency of \(150\ \mathrm{Hz}\) and the maximum user speed is given by \(v_d = c \frac{f_{\text{Doppler}}}{f_{\text{system}}} \approx 30\ \mathrm{m/s}\), where \(c\) denotes the speed of light. The temporal coherence time is approximately \(t_c = \frac{0.423}{f_{\text{Doppler}}} \approx 2.28\ \mathrm{ms}\). Assuming two samples per symbol, the resulting block duration is \(128\ \mu\mathrm{s}\), satisfying \(128\ \mu\mathrm{s} \ll 2.28\ \mathrm{ms}\), which validates the assumption of a constant channel within each transmission block. Under these conditions, the temporal correlation coefficient between adjacent blocks is approximately \(\rho \approx 0.9990\).

Fig.~\ref{fig:comparison} presents a comparison between \(\mathrm{NMSE}_{\mathrm{last}}\) and \(\mathrm{NMSE}_{\mathrm{coarse}}\), according to~(\ref{eq:comparison}), illustrating the advantage of exploiting temporal coherence in this scenario for LOS and NLOS scenarios. 

Figure~\ref{fig:efficiency} illustrates the spectral efficiency for both LOS and NLOS scenarios. We notice that, while keeping the code rate constant, the spectral efficiency depends on both the block length and the number of blocks that exploit temporal coherence. Due to the larger number of pilots required, a greater spectral efficiency gain is observed in the NLOS scenario. The negative values occur when the pilot length exceeds the block length. 

\begin{figure}
    \vspace{-1em}
\includegraphics[width=0.425\textwidth]{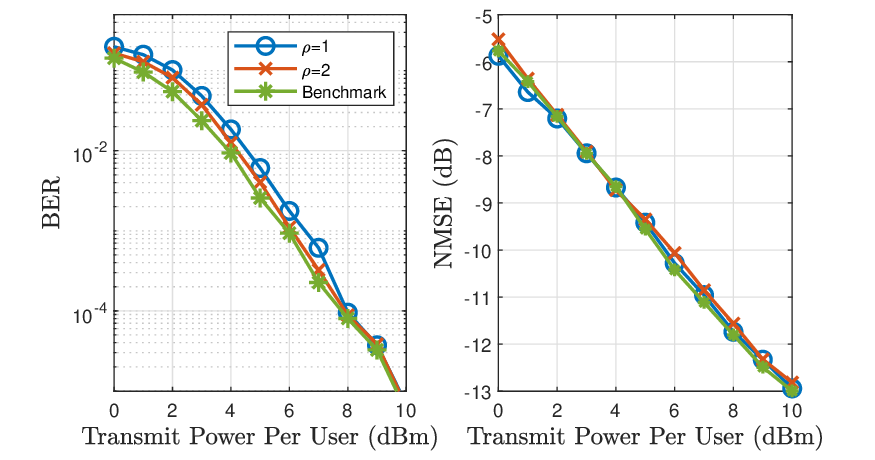}
  \vspace{-1em}
    \caption{LOS RIS-Assisted system with $N_p=N_p'=16$.}
    \label{fig:LOS}
    \vspace{-1em}
\end{figure}

\begin{figure}
    \vspace{-1em}
\includegraphics[width=0.425\textwidth]{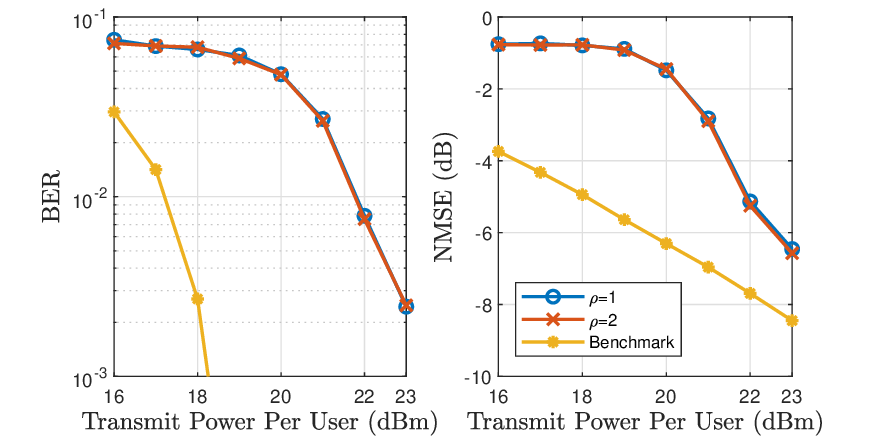}
  \vspace{-1em}
    \caption{NLOS RIS-Assisted system with $N_p = 96$  and  $N_p'' = 16$.}
    \label{fig:NLOS}
    \vspace{-1em}
\end{figure}

Fig.~\ref{fig:LOS} evaluates the performance of the proposed scheme in the LOS scenario for different numbers of iterative channel estimation steps, exploiting temporal coherence over $N_f = 20$ frames. The parameter $\rho$ denotes the number of iterations used to enhance the channel estimate. The results demonstrate that the proposed technique achieves comparable estimation performance, even when using uncoded pilot symbols.

In the NLOS scenario, shown in Fig.~\ref{fig:NLOS}, shorter coded pilot sequences are employed to improve spectral efficiency, which comes at the cost of a performance degradation. Nonetheless, the results indicate that as the SNR increases, the NMSE of the proposed scheme tends to approach that of the benchmark. It is worth noting that, due to the absence of a direct link, the proposed scheme requires a greater number of pilot symbols. Although the estimation performance remains inferior to the benchmark, the gap narrows at higher SNRs.

If the paper is accepted, we plan to include results with Beyond Diagonal and Holographic RIS architectures, which are expected to improve spatial resolution and wavefront control, potentially enhancing estimation accuracy and spectral efficiency under complex propagation conditions.
\vspace{-0.5em}

     \section{Conclusion}
\vspace{-0.25em}

This paper proposed a novel iterative detection, decoding, and channel estimation framework for RIS-assisted MIMO systems. Unlike conventional methods, the proposed scheme jointly exploits channel coding and encoded pilots to enhance estimation accuracy while significantly reducing the pilot overhead in both LOS and NLOS conditions. Furthermore, a channel tracking protocol based on temporal coherence was introduced, enabling improved estimation performance across successive blocks or a reduction in pilot symbols without substantial loss in accuracy. Simulation results validate the proposed approach, demonstrating substantial gains in NMSE and BER, particularly in NLOS scenarios, where conventional methods require a larger number of pilots.
\vspace{-0.5em}

%In this study, we proposed a novel iterative detection, decoding, and channel estimation scheme for RIS-assisted MIMO systems. Unlike existing approaches, our method exploits coding in the uplink to use parity bits for both decoding and channel estimation while employing encoded pilots to enhance performance. The proposed scheme significantly reduces the minimum number of pilots required for both LOS and NLOS scenarios, achieving superior performance in LOS scenarios. Numerical results show that our approach has large performance gains in terms of channel estimation NMSE and BER. \vspace{-0.25em}

     \bibliography{support/dyn_ris}

\begin{thebibliography}{10}
\providecommand{\url}[1]{#1}
\csname url@samestyle\endcsname
\providecommand{\newblock}{\relax}
\providecommand{\bibinfo}[2]{#2}
\providecommand{\BIBentrySTDinterwordspacing}{\spaceskip=0pt\relax}
\providecommand{\BIBentryALTinterwordstretchfactor}{4}
\providecommand{\BIBentryALTinterwordspacing}{\spaceskip=\fontdimen2\font plus
\BIBentryALTinterwordstretchfactor\fontdimen3\font minus \fontdimen4\font\relax}
\providecommand{\BIBforeignlanguage}[2]{{%
\expandafter\ifx\csname l@#1\endcsname\relax
\typeout{** WARNING: IEEEtran.bst: No hyphenation pattern has been}%
\typeout{** loaded for the language `#1'. Using the pattern for}%
\typeout{** the default language instead.}%
\else
\language=\csname l@#1\endcsname
\fi
#2}}
\providecommand{\BIBdecl}{\relax}
\BIBdecl

\bibitem{10767769}
N.~Li, R.~Xiong, Y.~Deng, F.~Xu, and H.~Deng, ``Semi-blind joint channel estimation and symbol detection for multi-{RIS}-aided {MIMO} systems,'' \emph{IEEE Wireless Communications Letters}, vol.~14, no.~3, pp. 586--590, 2025.

\bibitem{10439018}
M.~Fu, W.~Mei, and R.~Zhang, ``Multi-passive/active-irs enhanced wireless coverage: Deployment optimization and cost-performance trade-off,'' \emph{IEEE Transactions on Wireless Communications}, vol.~23, no.~8, pp. 9657--9671, 2024.

\bibitem{10818440}
H.~Khaleghi and A.~Haskou, ``Optimized channel estimation strategies for ris-aided communication,'' \emph{IEEE Communications Letters}, vol.~29, no.~3, pp. 453--456, 2025.

\bibitem{9854102}
Y.~Wei, M.-M. Zhao, A.~Liu, and M.-J. Zhao, ``Channel tracking and prediction for irs-aided wireless communications,'' \emph{IEEE Transactions on Wireless Communications}, vol.~22, no.~1, pp. 563--579, 2023.

\bibitem{9130088}
Z.~Wang, L.~Liu, and S.~Cui, ``Channel estimation for intelligent reflecting surface assisted multiuser communications: Framework, algorithms, and analysis,'' \emph{IEEE Transactions on Wireless Communications}, vol.~19, no.~10, pp. 6607--6620, 2020.

\bibitem{9839429}
H.~Guo and V.~K.~N. Lau, ``Uplink cascaded channel estimation for intelligent reflecting surface assisted multiuser miso systems,'' \emph{IEEE Transactions on Signal Processing}, vol.~70, pp. 3964--3977, 2022.

\bibitem{10614235}
A.~Agarwal, A.~Mishra, A.~Ray, and P.~Das, ``Sparse channel estimation in irs-assisted massive mimo cognitive radio systems,'' \emph{IEEE Transactions on Communications}, vol.~73, no.~1, pp. 200--215, 2025.

\bibitem{10484981}
H.~Fang, X.~Qiao, Y.~Zhang, L.~Yang, and H.~Zhu, ``On the performance of {RIS}-aided cell-free massive mimo systems under channel aging,'' \emph{IEEE Transactions on Vehicular Technology}, vol.~73, no.~9, pp. 12\,828--12\,841, 2024.

\bibitem{4357052}
M.-K. Oh, H.~M. Kwon, D.-J. Park, and Y.~H. Lee, ``Iterative channel estimation and {LDPC} decoding with encoded pilots,'' \emph{IEEE Transactions on Vehicular Technology}, vol.~57, no.~1, pp. 273--285, 2008.

\bibitem{10319806}
B.~Shi, Y.~Wu, P.~Yuan, D.~W.~K. Ng, X.-G. Xia, and W.~Zhang, ``Code-aided channel estimation in ldpc-coded mimo systems,'' \emph{IEEE Wireless Communications Letters}, vol.~13, no.~2, pp. 491--495, 2024.

\bibitem{9998527}
Z.~Zhang, L.~Dai, X.~Chen, C.~Liu, F.~Yang, R.~Schober, and H.~V. Poor, ``Active ris vs. passive ris: Which will prevail in 6g?'' \emph{IEEE Transactions on Communications}, vol.~71, no.~3, pp. 1707--1725, 2023.

\bibitem{10308579}
A.~Mishra, Y.~Mao, C.~D’Andrea, S.~Buzzi, and B.~Clerckx, ``Transmitter side beyond-diagonal reconfigurable intelligent surface for massive mimo networks,'' \emph{IEEE Wireless Communications Letters}, vol.~13, no.~2, pp. 352--356, 2024.

\bibitem{10158690}
J.~An, C.~Xu, D.~W.~K. Ng, G.~C. Alexandropoulos, C.~Huang, C.~Yuen, and L.~Hanzo, ``Stacked intelligent metasurfaces for efficient holographic mimo communications in 6g,'' \emph{IEEE Journal on Selected Areas in Communications}, vol.~41, no.~8, pp. 2380--2396, 2023.

\bibitem{mmimo}
R.~C. de~Lamare, ``Massive mimo systems: Signal processing challenges and future trends,'' \emph{URSI Radio Science Bulletin}, vol. 2013, no. 347, pp. 8--20, 2013.

\bibitem{wence}
W.~Zhang, H.~Ren, C.~Pan, M.~Chen, R.~C. de~Lamare, B.~Du, and J.~Dai, ``Large-scale antenna systems with ul/dl hardware mismatch: Achievable rates analysis and calibration,'' \emph{IEEE Transactions on Communications}, vol.~63, no.~4, pp. 1216--1229, 2015.

\bibitem{10747209}
R.~C.~G. Porto and R.~C. de~Lamare, ``Iterative detection and decoding for multiuser systems based on mmse refinements with active or passive ris,'' \emph{IEEE Wireless Communications Letters}, vol.~14, no.~1, pp. 208--212, 2025.

\bibitem{jidf}
R.~C. de~Lamare and R.~Sampaio-Neto, ``Adaptive reduced-rank processing based on joint and iterative interpolation, decimation, and filtering,'' \emph{IEEE Transactions on Signal Processing}, vol.~57, no.~7, pp. 2503--2514, 2009.

\bibitem{spa}
------, ``Minimum mean-squared error iterative successive parallel arbitrated decision feedback detectors for ds-cdma systems,'' \emph{IEEE Transactions on Communications}, vol.~56, no.~5, pp. 778--789, 2008.

\bibitem{mfsic}
P.~Li, R.~C. de~Lamare, and R.~Fa, ``Multiple feedback successive interference cancellation detection for multiuser mimo systems,'' \emph{IEEE Transactions on Wireless Communications}, vol.~10, no.~8, pp. 2434--2439, 2011.

\bibitem{dfcc}
P.~Li and R.~C. De~Lamare, ``Adaptive decision-feedback detection with constellation constraints for mimo systems,'' \emph{IEEE Transactions on Vehicular Technology}, vol.~61, no.~2, pp. 853--859, 2012.

\bibitem{mbdf}
R.~C. de~Lamare, ``Adaptive and iterative multi-branch mmse decision feedback detection algorithms for multi-antenna systems,'' \emph{IEEE Transactions on Wireless Communications}, vol.~12, no.~10, pp. 5294--5308, 2013.

\bibitem{bfidd}
A.~G.~D. Uchoa, C.~T. Healy, and R.~C. de~Lamare, ``Iterative detection and decoding algorithms for mimo systems in block-fading channels using ldpc codes,'' \emph{IEEE Transactions on Vehicular Technology}, vol.~65, no.~4, pp. 2735--2741, 2016.

\bibitem{did}
P.~Li and R.~C. de~Lamare, ``Distributed iterative detection with reduced message passing for networked mimo cellular systems,'' \emph{IEEE Transactions on Vehicular Technology}, vol.~63, no.~6, pp. 2947--2954, 2014.

\bibitem{rrber}
Y.~Cai, R.~C. de~Lamare, B.~Champagne, B.~Qin, and M.~Zhao, ``Adaptive reduced-rank receive processing based on minimum symbol-error-rate criterion for large-scale multiple-antenna systems,'' \emph{IEEE Transactions on Communications}, vol.~63, no.~11, pp. 4185--4201, 2015.

\bibitem{listmtc}
R.~B. Di~Renna and R.~C. de~Lamare, ``Iterative list detection and decoding for massive machine-type communications,'' \emph{IEEE Transactions on Communications}, vol.~68, no.~10, pp. 6276--6288, 2020.

\bibitem{msgamp}
R.~B.~D. Renna and R.~C. de~Lamare, ``Dynamic message scheduling based on activity-aware residual belief propagation for asynchronous mmtc,'' \emph{IEEE Wireless Communications Letters}, vol.~10, no.~6, pp. 1290--1294, 2021.

\bibitem{msgamp2}
R.~B. Di~Renna and R.~C. de~Lamare, ``Joint channel estimation, activity detection and data decoding based on dynamic message-scheduling strategies for mmtc,'' \emph{IEEE Transactions on Communications}, vol.~70, no.~4, pp. 2464--2479, 2022.

\bibitem{comp}
A.~B. L.~B. Fernandes, Z.~Shao, L.~T.~N. Landau, and R.~C. de~Lamare, ``Multiuser-mimo systems using comparator network-aided receivers with 1-bit quantization,'' \emph{IEEE Transactions on Communications}, vol.~71, no.~2, pp. 908--922, 2023.

\bibitem{vrce}
A.~Tang, J.-B. Wang, Y.~Pan, W.~Zhang, X.~Zhang, Y.~Chen, H.~Yu, and R.~C. de~Lamare, ``Joint visibility region and channel estimation for extremely large-scale mimo systems,'' \emph{IEEE Transactions on Communications}, vol.~72, no.~10, pp. 6087--6101, 2024.

\bibitem{llrap}
R.~B.~D. Renna and R.~C. de~Lamare, ``Iterative detection and decoding with log-likelihood ratio based access point selection for cell-free mimo systems,'' \emph{IEEE Transactions on Vehicular Technology}, vol.~73, no.~5, pp. 7418--7423, 2024.

\bibitem{iddocl}
T.~Ssettumba, S.~Mashdour, L.~T.~N. Landau, P.~B. da~Silva, and R.~C. de~Lamare, ``Iterative interference cancellation for clustered cell-free massive mimo networks,'' \emph{IEEE Wireless Communications Letters}, vol.~14, no.~2, pp. 509--513, 2025.

\bibitem{ldpc}
R.~Gallager, ``Low-density parity-check codes,'' \emph{IRE Transactions on Information Theory}, vol.~8, no.~1, pp. 21--28, 1962.

\bibitem{memd}
C.~T. Healy and R.~C. de~Lamare, ``Design of ldpc codes based on multipath emd strategies for progressive edge growth,'' \emph{IEEE Transactions on Communications}, vol.~64, no.~8, pp. 3208--3219, 2016.

\bibitem{vfap}
J.~Liu and R.~C. de~Lamare, ``Low-latency reweighted belief propagation decoding for ldpc codes,'' \emph{IEEE Communications Letters}, vol.~16, no.~10, pp. 1660--1663, 2012.

\bibitem{access2010further}
E.~U. T.~R. Access, ``Further advancements for {E-UTRA} physical layer aspects (rel. 9),'' \emph{European Telecommun. Standards Institute}, 2010.

\end{thebibliography}
\end{document}